# Quantitative phase imaging (QPI) through random diffusers using a diffractive optical network

*Yuhang Li*[1,2,3]    e-mail: yuhangli@ucla.edu

*Yi Luo*[1,2,3]    e-mail: yluo2016@ucla.edu

*Deniz Mengu*[1,2,3]    e-mail: denizmengu@ucla.edu

*Bijie Bai*[1,2,3]    e-mail: baibijie@ucla.edu

*Aydogan Ozcan*[1,2,3]*    e-mail: ozcan@ucla.edu

[1]Electrical and Computer Engineering Department, University of California, Los Angeles, California 90095, USA

[2]Bioengineering Department, University of California, Los Angeles, California 90095, USA

[3]California NanoSystems Institute (CNSI), University of California, Los Angeles, California 90095, USA

*Correspondence: Prof. Aydogan Ozcan  E-mail: ozcan@ucla.edu

Address: 420 Westwood Plaza, Engr. IV 68-119, UCLA, Los Angeles, CA 90095, USA

Tel: +1(310)825-0915

Fax: +1(310)206-4685

## Abstract

Quantitative phase imaging (QPI) is a label-free computational imaging technique used in various fields, including biology and medical research. Modern QPI systems typically rely on digital processing using iterative algorithms for phase retrieval and image reconstruction. Here, we report a diffractive optical network trained to convert the phase information of input objects positioned behind random diffusers into intensity variations at the output plane, all-optically performing phase recovery and quantitative imaging of phase objects completely hidden by unknown, random phase diffusers. This QPI diffractive network is composed of successive diffractive layers, axially spanning in total ~$70\lambda$, where $\lambda$ is the illumination wavelength; unlike existing digital image reconstruction and phase retrieval methods, it forms an all-optical processor that does not require external power beyond the illumination beam to complete its QPI reconstruction at the speed of light propagation. This all-optical diffractive processor can provide a low-power, high frame rate and compact alternative for quantitative imaging of phase objects through random, unknown diffusers and can operate at different parts of the electromagnetic spectrum for various applications in biomedical imaging and sensing. The presented QPI diffractive designs can be integrated onto the active area of standard CCD/CMOS-based image sensors to convert an existing optical microscope into a diffractive QPI microscope, performing phase recovery and image reconstruction on a chip through light diffraction within passive structured layers.

## Introduction

Imaging weakly scattering phase objects, such as cells, has been an active research area for decades, with various solutions reported[1–6] for applications in different fields, including biomedical sciences. One common approach is the use of chemical stains[7] or fluorescent tags[8] to bring contrast to such weakly scattering microscopic features of objects, but these methods require complex sample preparation steps, involving the use of exogenous labeling agents, which might interfere with the normal physiological processes of specimens. Differential interference contrast (DIC) microscopy is another commonly used method, which can rapidly image optical path length changes in unstained samples, revealing the qualitative phase information[9,10]; however, it lacks a quantitative measurement of the optical phase distribution. Addressing the need for quantifying the phase shift information of objects, quantitative phase imaging (QPI) has become a powerful and widely used tool for non-invasively imaging transparent specimens with high sensitivity and resolution[11]. Over the last decades, various digital QPI techniques have been developed, such as Fourier Phase Microscopy (FPM)[12], Digital Holographic Microscopy (DHM)[3,13,14], Diffraction Phase Microscopy (DPM)[15], Spatial Light Interference Microscopy (SLIM),[16] among many others. Traditional QPI systems often require relatively large-scale computational resources for image reconstruction and phase retrieval algorithms, which are time-consuming, partially hindering the frame rate of these computational imaging systems. With the development and widespread use of deep learning methods, recent works have also involved deep neural networks in QPI, significantly advancing the image reconstruction speed and the spatiotemporal throughput, also improving the image quality by levering machine learning and GPU-based computing[6,17–28].

Recent research also presented an all-optical phase recovery and image reconstruction method for QPI using diffractive deep neural networks ($D^2NN$), enabling computer-free image reconstruction of phase objects at the speed of light propagation through thin diffractive layers[29]. A diffractive network is an all-optical machine learning platform that computes a given task using light diffraction through successive (passive) transmissive layers[30–35], where each diffractive layer typically consists of tens of thousands of diffractive units to modulate the phase and/or amplitude of the incident light. Deep learning techniques, such as error backpropagation and stochastic gradient descent, are used to optimize each layer's modulation values (e.g., transmission coefficients), mapping complex-valued input fields containing the optical information of interest onto desired output fields[36–38].

In this work, we report the design of diffractive optical networks for phase recovery and quantitative phase imaging through random unknown diffusers. Unlike some of the earlier work[29,38], the presented QPI diffractive network can convert the phase information of an input sample into a quantitative output intensity distribution even in the presence of unknown, random phase diffusers, all-optically revealing the quantitative phase images of the samples that are completely covered by random diffusers. This diffractive network, after its training, generalizes to all-optically perform QPI through unknown random diffusers never seen before, without the need for a digital image reconstruction algorithm. It has a compact axial thickness of ~$70\lambda$, and does not require any computing power except for the illumination light. The QPI $D^2NN$ designs reported in this work can potentially be integrated with existing CCD/CMOS image sensors by fabricating the resulting thin diffractive layers on top of the active area of an image sensor array. Such an on-chip integrated D²NN can be placed at the image plane of a standard microscope to convert it into a diffractive QPI microscope. This diffractive computing framework for phase retrieval and QPI through random unknown diffusers can potentially advance label-free microscopy and sensing applications in biomedical sciences, among other fields.

## Results

**Design of a diffractive optical network for QPI through random diffusers**

Figure 1(a) illustrates the schematic of a 4-layer QPI D²NN trained to all-optically recover the phase information of an input phase object through unknown random phase diffusers. To train this QPI diffractive network, phase-only objects with unit amplitude were randomly selected from the MNIST dataset and placed at $53.3\lambda$ in front of randomly generated phase diffusers (see the Materials and methods section). The QPI diffractive network designed here was composed of four successive diffractive layers with an axial distance of $2.67\lambda$ between them, and the distance between the random diffuser plane and the first diffractive layer was also $2.67\lambda$. The output image plane was designed to be $9.3\lambda$ away from the last diffractive layer, as shown in Fig .1(b).

We introduced multiple random diffusers in each training epoch to build the generalization capability for the diffractive layers to quantitatively image phase objects distorted by new random diffusers. We used the correlation length ($L$) to characterize random diffusers in terms of their effective grain size (see the Materials and methods section), and all the random diffusers used in training and blind testing were assumed to have the same correlation length ($L = L_{train} = L_{test}$), modeled as thin phase masks (Fig. 1(b)). During the training phase, handwritten digit samples were randomly selected from the MNIST dataset, and fed to the diffractive network, propagating through the corresponding random diffuser and the successive diffractive layers to form the intensity profiles at the output plane. The phase values of the diffractive features at each layer were adjusted through error backpropagation by minimizing the mean square error (MSE) between the target QPI images and the normalized output intensity profiles (see the Materials and methods section). One epoch was completed when all the 55,000 images in the MNIST dataset were used, and the training stopped after 200 epochs when the QPI D²NN used/saw $N = 200n$ different random diffusers, where $n$ is the number of diffusers used in each epoch (e.g., $n = 20$ and $N = 4000$). After the training, the converged QPI diffractive networks were numerically tested by imaging unknown phase objects through new, unseen diffusers, as shown in Fig. 1(b-c).

Without loss of generality, all the QPI diffractive networks reported in this paper were designed with unit magnification, i.e., the output intensity features have the same scale as the input phase features; this is not a limitation since the thin QPI D$^2$NN design (spanning ~70$\lambda$ in thickness) can be placed at the magnified image plane of a QPI microscope, by fabricating and integrating it on top of the active area or the protective glass of the CMOS/CCD-based imager chip. Since the output optical intensity at the back-end of the diffractive network depends on the power of the illumination source, the diffraction efficiency of transmissive layers and the quantum efficiency of the image sensor array, we defined a reference region at the output plane, within which the mean signal intensity was calculated to normalize the raw output intensity of the QPI D$^2$NN (see the Materials and methods section). After this normalization, the resulting output intensity, denoted as $I_{QPI}(x, y)$ $[rad]$, was used as the final quantitative phase image. This makes the QPI D$^2$NN output images *independent* of external factors such as the illumination beam intensity or the quantum efficiency of the image sensor used as part of the microscope design, helping us quantitatively map the phase information of the samples behind unknown diffusers into intensity signals.

**All-optical phase recovery through random diffusers using QPI diffractive networks**

To demonstrate the all-optical phase recovery through random unknown diffusers using QPI diffractive networks, we used the samples from the MNIST dataset as phase-only input objects and trained a 4-layer network with $n = 20$ random diffusers in each epoch, i.e., $N = 4000$ random diffusers were used in total. To test the performance of the trained QPI D$^2$NN model, we first used new hand-written digits from the test set that were never used during the training stage; these test objects were individually distorted by $n = 20$ random diffusers used in the last training epoch (termed as *known* diffusers) as well as some newly generated diffusers that were never used during the training (termed as *new* diffusers), as shown in Fig. 2(a). The resulting output images of the QPI diffractive network reveal its generalization performance for all-optical phase recovery and quantitative phase imaging of new test objects through new random phase diffusers that were never seen before.

We further tested the same QPI D²NN, which was trained only with MNIST handwritten digits, using binary phase gratings to quantify the smallest resolvable linewidth and the phase sensitivity of the all-optical image reconstructions through *new* random diffusers; see Fig. 3. In this analysis, we varied the grating linewidth while keeping the binary phase contrast as $0 - \pi$; each phase grating at the input plane was completely hidden behind random unknown phase diffusers as before. Our numerical results reported in Fig. 3(a) show that $0 - \pi$ phase encoded gratings with a linewidth of ~$9.6\lambda$ were resolvable by our QPI D²NN regardless of the grating direction and the random phase diffuser used. Despite being trained using only handwritten digits with relatively poor resolution, the QPI diffractive network was able to quantitatively reconstruct these phase gratings through unknown random diffusers, indicating that our diffractive model was successful in approximating a general-purpose quantitative phase imager. Its resolution can be further improved by using training images that contain higher resolution, sharper features.

We also tested the same QPI D²NN network to image distortion-free gratings by removing the random phase diffusers in Fig. 1(b) while keeping all other components unchanged; this scheme is *against* our training which always used a random phase diffuser behind the input plane. Despite deviating from its training configuration, the QPI D²NN showed better image reconstruction quality when the random diffusers were removed, further demonstrating that the diffractive network design converged to a general-purpose quantitative phase imager, converting the phase information at the input plane into quantitative intensity patterns at its output, *with* and *without* the presence of random phase diffusers.

The input phase contrast is another factor affecting the resolution achieved by our QPI diffractive network design. To shed more light on this, we numerically evaluated our QPI D²NN on binary phase gratings at varying levels of input phase contrast (Fig. 3(b)); through this analysis, we found out that the input phase gratings with a linewidth of $9.6\lambda$ remained resolvable even when the input phase contrast was reduced to $0.25\pi$. We also performed a similar phase contrast analysis using the test samples from the MNIST dataset to further examine the impact of the input phase contrast over the quality of the QPI reconstructions created by the diffractive optical network trained with $\alpha_{train} = \alpha = 1$, as shown in Fig. 4(a), where $\alpha$ denotes the phase range $[0, \alpha \cdot \pi]$ used for the training images. During the blind testing, the input phase contrast parameter ($\alpha_{test}$) was varied from 0.1 to 1.25, and the reconstructed D²NN images at the output plane were quantified using the Pearson Correlation

Coefficient (PCC) and the percent phase error (see the Materials and methods section). Figures 4(b)-(c) illustrate the mean and the standard deviation of the resulting PCC and the percent phase error values as a function of $\alpha_{test}$, both of which peak at $\alpha_{test} = 0.75$ rather than $\alpha_{test} = 1$; this is due to the continuous distribution of the phase values of the training images spanning $[0, \alpha_{train} \cdot \pi]$. There is a performance drop as $\alpha_{test}$ decreased to 0.1, suggesting that the trained QPI diffractive network has difficulty separating the foreground and background for smaller phase contrast input objects that are hidden behind random unknown diffusers. On the other hand, when $\alpha_{test}$ increased to 1.25, beyond its training range $\alpha_{train} = 1$, the reconstructed image quality was still acceptable, although some performance degradation appears (Figs. 4(b-d)), demonstrating the capability of the QPI D²NN to generalize to input objects exceeding the phase contrast range used during the training stage.

All these analyses reported above were performed using phase-only input test objects that were completely hidden behind random unknown phase diffusers with $L_{test} = 14\lambda$. Next, we removed the random phase diffusers in Fig. 4(a) and tested the phase contrast performance of the same QPI D²NN to quantitatively image distortion-free phase-only objects; the results of this analysis are plotted in Figs. 4(b)-(c) (red curves) as $\alpha_{test}$ varied from 0.1 to 1.25. As expected, Figs. 4(b)-(d) reveal that the performance of the same trained diffractive QPI network ($\alpha_{train} = 1$) was much better when the random diffusers were removed, further supporting that the D²NN converged to a general-purpose quantitative phase imager, which is not only able to perform phase-to-intensity transformations but is also resilient to structural distortions caused by random, unknown phase diffusers hiding the input phase objects.

**Impact of the number of diffractive layers**

Through both theoretical and empirical evidence, it was demonstrated that deeper diffractive optical networks compute an arbitrary complex-valued linear transformation with lower approximation errors, and such deeper diffractive architectures exhibit higher generalization capacity for all-optical statistical inference tasks[36,37,39,40]. Similarly, we also analyzed the impact of the number ($K$) of trainable diffractive layers on the all-optical phase recovery and QPI performance for imaging phase-only objects through random unknown diffusers. Figure 5(a) reports the output images of the QPI diffractive networks through *known*, *new* and *no* diffusers, where the known diffusers refer to the random diffusers used in the last training epoch, and the new diffusers are the newly generated random diffusers, never seen before. Figures 5(b)-(c) compare the average PCC values and the absolute phase errors for the phase imaging of unknown test objects using diffractive QPI networks designed with different $K$, showing that 2-layer networks had relatively low PCC values and high phase error, and the imaging performance improved as we increased the number of diffractive layers, $K$. This can also be visualized in Figure 5(a), where the QPI results of the 2-layer diffractive design are blurry with low contrast compared to the results of the 6-layer diffractive design. Our results further reveal that, with the additional trainable diffractive layers available, the average PCC values in all three cases (i.e., known, new and no diffusers) increase, while the absolute phase errors decrease.

**Impact of the diffuser correlation length (*L*)**

We also investigated how the diffusers' correlation length affects the imaging quality. For this analysis, we designed three different QPI diffractive networks, each trained using random diffusers with a correlation length ($L_{train}$) of $10\lambda, 14\lambda$ and $17\lambda$ (see the Materials and methods section); each one of the resulting QPI D²NN was blindly tested with random new phase diffusers with the same correlation length $L_{test} = L_{train}$. Figure 6(a) visualizes the output images of these three QPI networks, revealing that the diffractive networks trained and tested with larger correlation length diffusers more accurately reflect the original phase distribution at the input, which aligns with the fact that the phase diffusers with larger correlation lengths introduce weaker distortions to the input objects. Figures 6(b)-(c) plot the PCC values and the phase errors of these three QPI D²NN models for known, *forgotten*, new and no diffusers, where the "forgotten" diffusers refer to the diffusers used during the training stage before the last epoch, i.e., the random diffusers used from the 1st to the 199th epoch of our training. We observe that the QPI performance through the known diffusers used in the final (200th) epoch of the training is slightly better than imaging through forgotten diffusers or new diffusers, which is expected due to the partial "memory" of the diffractive QPI network. Another important finding is that the all-optical phase recovery and QPI performance of these trained diffractive networks to image test objects through new random diffusers is comparable to imaging through forgotten diffusers. Stated differently, from the perspective of the QPI D²NN, a new random phase diffuser is statistically identical to a forgotten phase diffuser that was used during the earlier epochs of the training; in fact, this feature can be considered a signature of successful training of a diffractive imager network to see through random diffusers.

**The trade-off between the QPI quality and the output diffraction efficiency**

Two factors mainly influence the output power efficiency of the presented QPI networks: the diffraction efficiency of the transmissive layers and the material absorption. In this study, we assumed that the absorption of the optical material of the diffractive layers is negligible for the operating wavelength of interest; this is a valid assumption for most materials in the visible band (e.g., glass and polymers) since the entire axial thickness of a QPI $D^2NN$ design is <100$\lambda$. To control and accordingly enhance the output power efficiency of QPI diffractive networks, an additional loss function was introduced, which balanced the trade-off between the QPI performance and the diffraction efficiency (see the Materials and methods section). In Fig. 7, we present two QPI $D^2NN$ designs with $K$=4 and 8 diffractive layers, both sharing the same parameters as the setup shown in Fig. 1. Their QPI performance through new diffusers and no diffusers are also plotted in Fig. 7. For the 4-layer QPI $D^2NN$ architecture, the previously presented diffractive model that was designed *without* a power-efficiency penalty (Fig. 2) achieved an output power efficiency of ~0.5% and a PCC value of 0.885 for imaging input objects through new random diffusers. By introducing the additional power-efficiency loss term during training, the same QPI $D^2NN$ architecture with $K$=4 achieved an increased output power efficiency of ~1.86% while maintaining a good output image quality with a PCC of 0.831. Compared to the original QPI $D^2NN$ design that solely focused on the output image quality, the newly trained diffractive network, which took into account both the image quality and output power efficiency, improved the diffraction efficiency by ~3-fold, with only a minor compromise on the output image quality. As shown in Fig. 7, for the QPI $D^2NN$ models with $K$=8 diffractive layers, the output image quality was further improved compared to the 4-layer designs at the same output efficiency performance. In general, a deeper $D^2NN$ architecture, such as the 8-layer model, can achieve a better trade-off between the output diffraction efficiency and the QPI performance compared to shallow $D^2NN$ models with fewer layers.

**Discussion**

As demonstrated in our numerical results, a QPI $D^2NN$ trained with the MNIST dataset can all-optically recover the phase information of unknown test objects completely covered by random unknown diffusers. By using the mean intensity value surrounding the QPI signal area as a normalization term, the QPI network becomes invariant to changes in the input beam intensity or the power efficiency of the system, and the resulting normalized intensity profiles quantify the phase distribution of the input objects distorted by random diffusers. Since these QPI diffractive networks only consist of passive diffractive layers, they perform phase recovery and quantitative phase imaging without needing an external power source except for the illumination light. Although the training process takes relatively long (e.g., ~72 hours), it's a one-time effort; after this one-time training and the fabrication of the resulting diffractive layers, the quantitative phase imaging of specimen hidden by unknown random phase diffusers can be performed at the speed of light propagation through a thin optical volume that axially spans <100λ.

Although we considered here random diffusers as single-layer thin phase elements, which is a common assumption in various applications[41–44], an extension of this QPI D²NN concept for imaging through volumetric diffusers is left as future work, which might find broader applications[45,46]. For example, the design of a hybrid system, which jointly trains a front-end diffractive network and a back-end electronic neural network[31,33,47] may be used to boost the performance of QPI through more complicated volumetric random diffusers. Finally, our results and methods can be extended to operate at various parts of the electromagnetic spectrum, including the visible and infrared wavelengths.

## Materials and methods

**The design of the random phase diffusers**

We modeled a random phase diffuser as a phase-only mask, whose complex transmission coefficient $t_D(x,y)$ is defined by the refractive index difference between the air and the diffuser material ($\Delta n \approx 0.74$) and a random heightmap $D(x,y)$ at the diffuser plane, i.e.,

$$t_D(x,y) = exp\left(j\frac{2\pi\Delta n}{\lambda}D(x,y)\right) \quad (1)$$

where $j = \sqrt{-1}$. The random height map $D(x,y)$ is defined as

$$D(x,y) = W(x,y) * K(\sigma) \quad (2)$$

where $W(x,y)$ follows a normal distribution with a mean μ and a standard deviation $\sigma_0$, i.e.

$$W(x,y) \sim \mathcal{N}(\mu, \sigma_0^2) \quad (3)$$

$K(\sigma)$ is a zero-mean Gaussian smoothing kernel with a standard deviation of σ, and ' ∗ ' denotes the 2D convolution operation. The phase-autocorrelation function $R_d(x, y)$ of a random phase diffuser is related to the correlation length $L$ as:

$$R_d(x, y) = \exp(-\pi(x^2 + y^2)/L^2) \qquad (4)$$

By numerically fitting the function $\exp(-\pi(x^2 + y^2)/L^2)$ to $R_d(x, y)$, we can statistically get the correlation length $L$ of randomly generated diffusers. In this work, for μ = 25λ, $\sigma_0$ = 8λ and σ = 7λ, we calculated the average correlation length as $L \sim 14\lambda$ based on 2000 randomly generated phase diffusers. We accordingly modified the σ values to generate the corresponding random phase diffusers for the other correlation lengths $L$ used in this work.

**Optical forward model of the QPI D²NN**

Free space propagation in air between the diffractive layers was formulated using the Rayleigh-Sommerfeld equation. The propagation can be modeled as a shift-invariant linear system with the impulse response:

$$w(x, y, z) = \frac{z}{r^2}\left(\frac{1}{2\pi r^2} + \frac{1}{j\lambda}\right) exp\left(\frac{j2\pi r}{\lambda}\right) \qquad (5)$$

where $r = \sqrt{x^2 + y^2 + z^2}$ and $n = 1$ for air. Considering a plane wave that is incident at a phase-modulated object $h(x, y, z = 0)$ positioned at $z = 0$, we formulated the distorted image right after the random phase diffuser located at $z_0$ as:

$$u_0(x, y, z_0) = t_D(x, y) \cdot [h(x, y, 0) * w(x, y, z_0)] \qquad (6)$$

This distorted field is used as the input field of subsequent diffractive layers. The diffractive layers were modeled as thin phase elements. Consequently, the transmission coefficient of the layer $m$ located at $z = z_m$ can be formulated as:

$$t_m = \exp(j\phi(x, y, z_m)) \qquad (7)$$

The optical field $u_m(x, y, z_m)$ right after the $m^{th}$ diffractive layer at $z = z_m$ can be written as:

$$u_m(x, y, z_m) = t_m(x, y, z_m) \cdot [u_{m-1}(x, y, z_{m-1}) * w(x, y, \Delta z_m)] \qquad (8)$$

where $\Delta z_m = z_m - z_{m-1}$ is the axial distance between two successive diffractive layers, which was selected as 2.67λ throughout this paper. After being modulated by all the $K$ diffractive layers, the optical field was collected at an output plane which was $\Delta z_d = 9.3\lambda$ away from the last diffractive layer. The intensity of this optical field is used as the raw output of the QPI D²NN:

$$I_{raw}(x,y) = |u_M * w(x,y,\Delta z_d)|^2 \qquad (9)$$

**The design of QPI diffractive networks**

During the training process of QPI D²NNs, we sampled the 2D space with a grid of $0.4\lambda$, which is also the size of each diffractive feature on the diffractive layer. A coherent light was assumed as the illumination source for the diffractive neural networks with a wavelength of $\lambda \approx 0.75$mm. As for the physical layout of the QPI D²NN, the input field-of-view (FOV) was set to be $96\lambda \times 96\lambda$, which corresponds to $240 \times 240$ pixels defining the phase distribution of the input objects. Handwritten digits from the MNIST training dataset were first normalized to the range $[0, 1]$ and bilinearly interpolated from $28 \times 28$ pixels to $14 \times 14$ ($\phi_{target}$). The resulting images were up-sampled to $168 \times 168$ using 'nearest' mode, then padded with zeros to $240 \times 240$ pixels ($\phi_i$), matching the size of the input FOV. Stated differently, without loss of generality, we defined an object-free region with a constant transmission coefficient of 1 to surround the samples of interest. The values of the padded images $\phi_i(x,y)$ were used to define the input phase values, and the amplitude at each pixel was taken as 1. Another parameter ($\alpha$) was introduced to control the range of the input phase; accordingly, the complex amplitude at the input FOV can be expressed as $input = e^{j\alpha\pi\phi_i}$ with a size of $240 \times 240$ pixels, and the target (ground truth) output intensity is $I_{target} = \alpha\pi\phi_{target}$ with a size of $14 \times 14$ pixels.

The physical size of each diffractive layer was also set to be $96\lambda \times 96\lambda$, i.e., each diffractive layer contained $240 \times 240$ trainable diffractive features, which only modulated the phase of the incident light field. The axial distances between the input phase object and the random diffuser, the diffuser and first diffractive layer, two successive diffractive layers, and the last diffractive layer and the output plane were set to be $53.3\lambda, 2.67\lambda, 2.67\lambda$ and $9.3\lambda$, respectively. The size of the signal area at the output plane, including the reference region, was set to be $69.6\lambda \times 69.6\lambda$ ($174 \times 174$ pixels), in which we cropped the central $67.2\lambda \times 67.2\lambda$ ($168 \times 168$ pixels) region as the QPI signal area and the edge region extending (in both directions on $x$ and $y$ axes) by 3 pixels was set as the *reference* region. According to our forward model, the QPI signal $I_{QPI}(x,y)$ can be written as:

$$I_{QPI}(x,y) = \frac{I_{raw}(x,y)}{Ref} \qquad (10)$$

where $Ref$ is the mean background intensity value within the reference region at the output plane, and $I_{QPI}(x,y)$ indicates the quantitative phase image in radians. We further cropped the central $168 \times 168$ pixels of $I_{QPI}$ and binned every $12 \times 12$ pixels to one pixel by averaging such that $I_{QPI}$ had a final size of $14 \times 14$ pixels representing the input object phase in radians.

During the training, $n$ uniquely different phase diffusers were randomly generated at each epoch. In each training iteration, a batch of $B = 10$ different objects from the MNIST handwritten digit dataset were sampled randomly; each input object in a batch was numerically duplicated $n$ times and separately perturbed by a set of $n$ randomly selected diffusers. Therefore, $B \times n$ different optical fields were obtained, and these distorted fields were individually forward propagated through the same state of the diffractive network. Therefore, we got $B \times n$ different normalized intensity patterns at the output plane ($I_{QPI\_1}, \ldots, I_{QPI\_Bn}$), which were used for the mean square error (MSE)-based training loss function calculation:

$$Loss = \frac{\frac{1}{N_{QPI}} \sum_{i=1}^{Bn} \sum_{x,y} |I_{target}(x,y) - I_{QPI\_i}(x,y)|^2}{Bn} \qquad (11)$$

where $N_{QPI} = 14 \times 14$.

Pearson Correlation Coefficient (PCC) was used to evaluate the linear correlation between the output QPI image $I_{QPI}(x,y)$ and the target $I_{target}(x,y)$, which can be expressed as:

$$PCC = \frac{\sum (I_{QPI}(x,y) - \overline{I_{QPI}}) \cdot (I_{target}(x,y) - \overline{I_{target}})}{\sqrt{\sum (I_{QPI}(x,y) - \overline{I_{QPI}})^2 \cdot (I_{target}(x,y) - \overline{I_{target}})^2}} \qquad (12)$$

We also calculated the absolute phase error to assess the phase recovery performance of a QPI D²NN:

$$phase\ error = \frac{1}{N_{QPI}} \sum_{x,y} |I_{target}(x,y) - I_{QPI}(x,y)| \qquad (13)$$

while the percent phase error is:

$$phase\ error\% = \frac{1}{N_{QPI}} \sum_{x,y} \frac{|I_{target}(x,y) - I_{QPI}(x,y)|}{I_{target}(x,y)} \qquad (14)$$

In analyzing the impact of reduced input phase contrast on the QPI performance, we trained a QPI D²NN using the MNIST dataset and tested it with binary gratings and handwritten digits. We binarized the MNIST samples by setting a threshold of 0.5 during the testing stage. In the exploration of the trade-off between the QPI performance and the output diffraction efficiency, we calculated the power-efficiency $E(I_{raw})$ of the QPI D²NN as:

$$E(I_{raw}) = \frac{\sum I_{raw}(x,y)}{\sum |input(x,y)|} = \frac{\sum I_{raw}(x,y)}{240^2} \quad (15)$$

and the corresponding diffraction efficiency penalty was calculated as follows:

$$Loss_{eff}(I_{raw}) = \max\{0, E_{target} - E(I_{raw})\} \quad (16)$$

where $E_{target}$ was the target power-efficiency, which varied from 0 to 0.03 for the models presented in Fig. 7; the diffractive model presented in Fig. 2 was trained without any diffraction efficiency penalty. Based on these definitions, the total loss function that included the power-efficiency penalty can be rewritten as:

$$Loss = \frac{\frac{1}{N_{QPI}} \sum_{i=1}^{Bn} \sum_{x,y} |I_{target}(x,y) - I_{QPI\_i}(x,y)|^2}{Bn} + \frac{\sum_{i=1}^{Bn} \max\{0, E_{target} - E(I_{raw\_i})\}}{Bn} \quad (17)$$

**Digital implementation**

The QPI diffractive neural networks were trained using Python (v3.6.13) and PyTorch (v1.11, Meta AI) with a GeForce GTX 1080 Ti graphical processing unit (GPU, Nvidia Corp.), an Intel® Core™ i7-7700K central processing unit (CPU, Intel Corp.) and 64 GB of RAM, running the Windows 10 operating system (Microsoft Corp.). The calculated loss values were backpropagated to update the diffractive layer transmission values using the Adam optimizer[48] with a decaying learning rate of $0.99^{epoch} \times 10^{-3}$, where $epoch$ refers to the current epoch number. Training a typical QPI D²NN model takes ~72 h to complete with 200 epochs and $n = 20$ diffusers per epoch.

**Figures and Figure Captions**

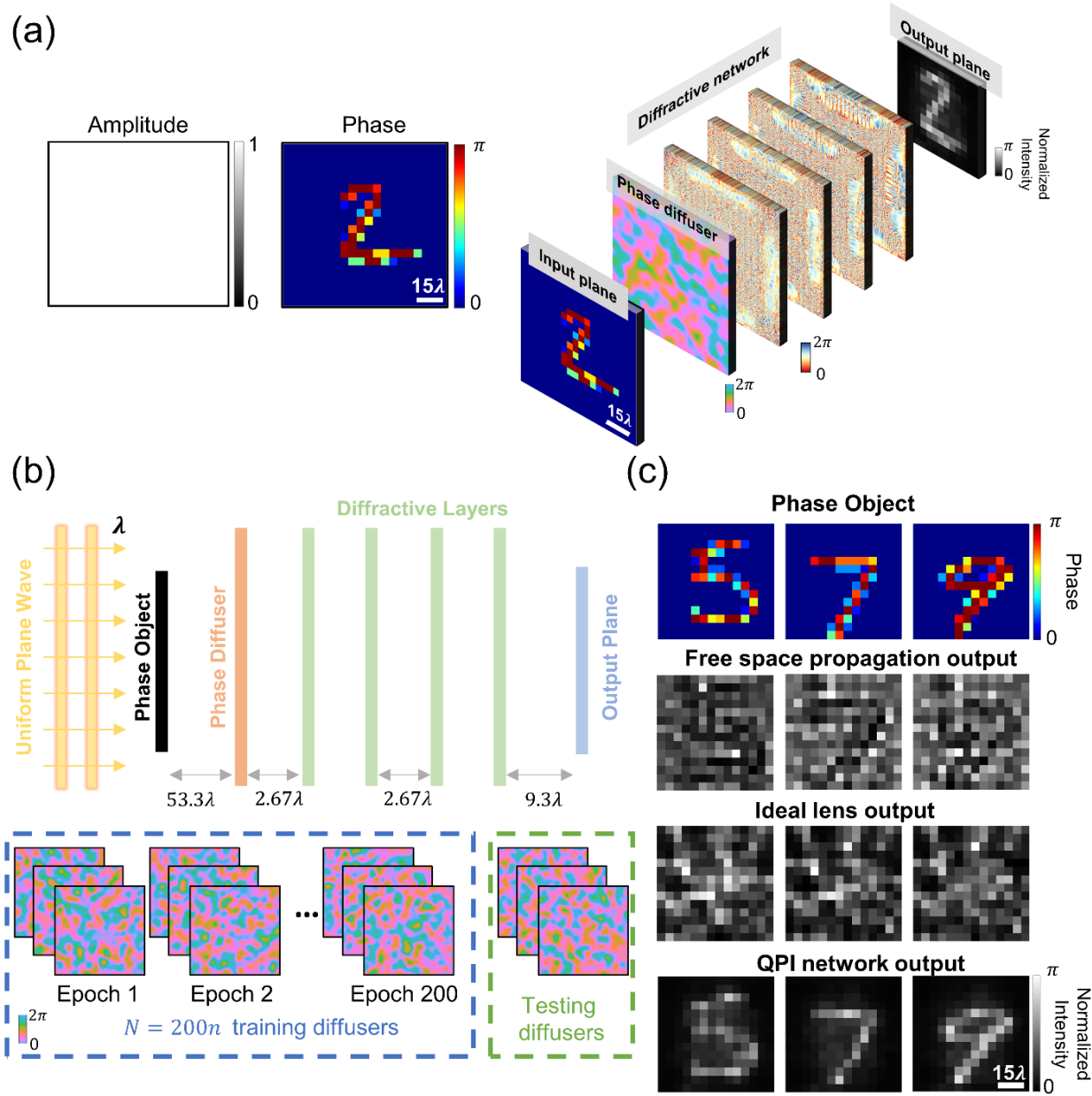

**Figure 1. All-optical phase recovery and quantitative phase imaging through random unknown diffusers using a D²NN. (a)** The schematic drawing of the presented QPI D²NN, converting the phase information of an input object behind a random phase diffuser into a normalized intensity image, which reveals the QPI information in radians without the use of a digital image reconstruction algorithm. **(b)** Optical layout and the training schematic of the presented QPI diffractive networks. **(c)** Sample images showing the image distortion generated by random phase diffusers with $L = 14\lambda$. Top: input phase objects. Second row: free space propagation (FSP) of the input objects through the diffusers, without the diffractive layers. Third row: the input objects imaged by an aberration-free lens through the random diffuser. Fourth row: the QPI D²NN output.

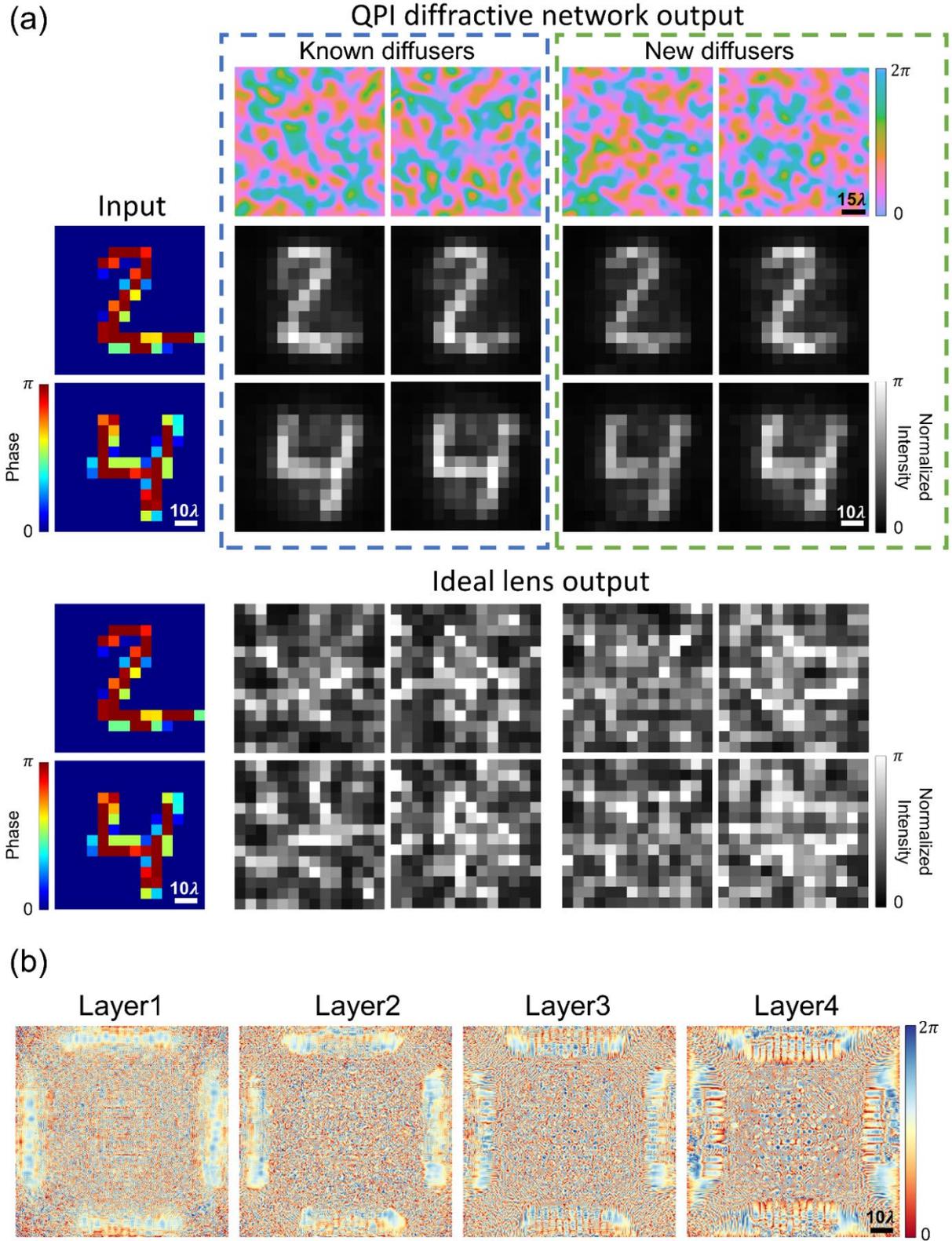

**Figure 2. Simulation results of the 4-layer QPI diffractive network for all-optical phase recovery through random diffusers. (a)** QPI D²NN phase recovery results through known and new random diffusers for phase-encoded handwritten digits '2' and '4' **(b)** The phase profiles of the trained diffractive layers of the QPI D²NN. $L = L_{train} = L_{test} = 14\lambda$.

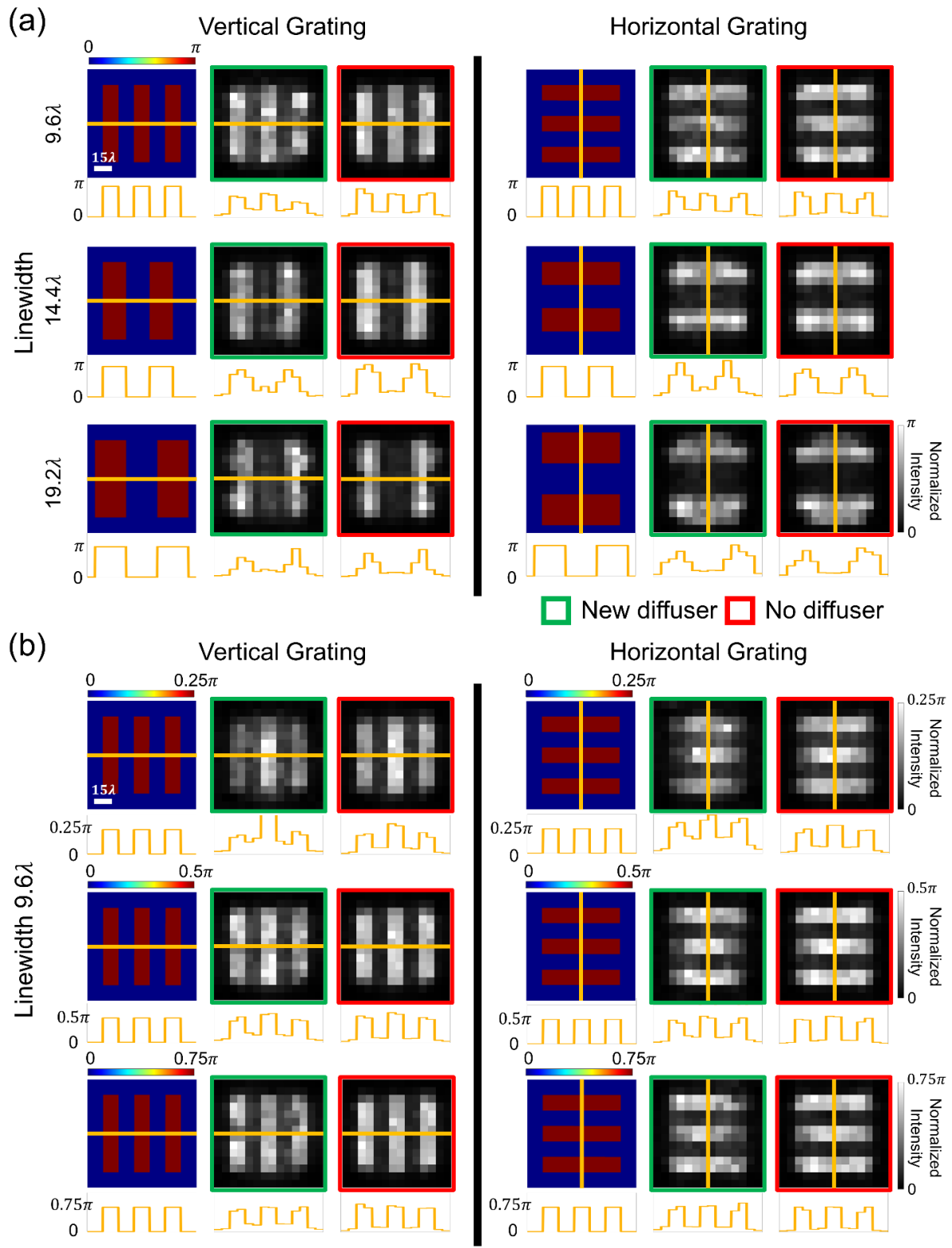

**Figure 3**. **Spatial resolution and phase sensitivity analyses of a diffractive QPI network, imaging through random diffusers.** **(a)** Input phase and the corresponding output QPI D²NN signal for binary $(0 - \pi)$ phase-encoded gratings with different linewidths. **(b)** Input phase and the corresponding output

signal of the QPI D²NN for binary phase-encoded gratings with different phase contrast values. $L = L_{train} = L_{test} = 14\lambda$.

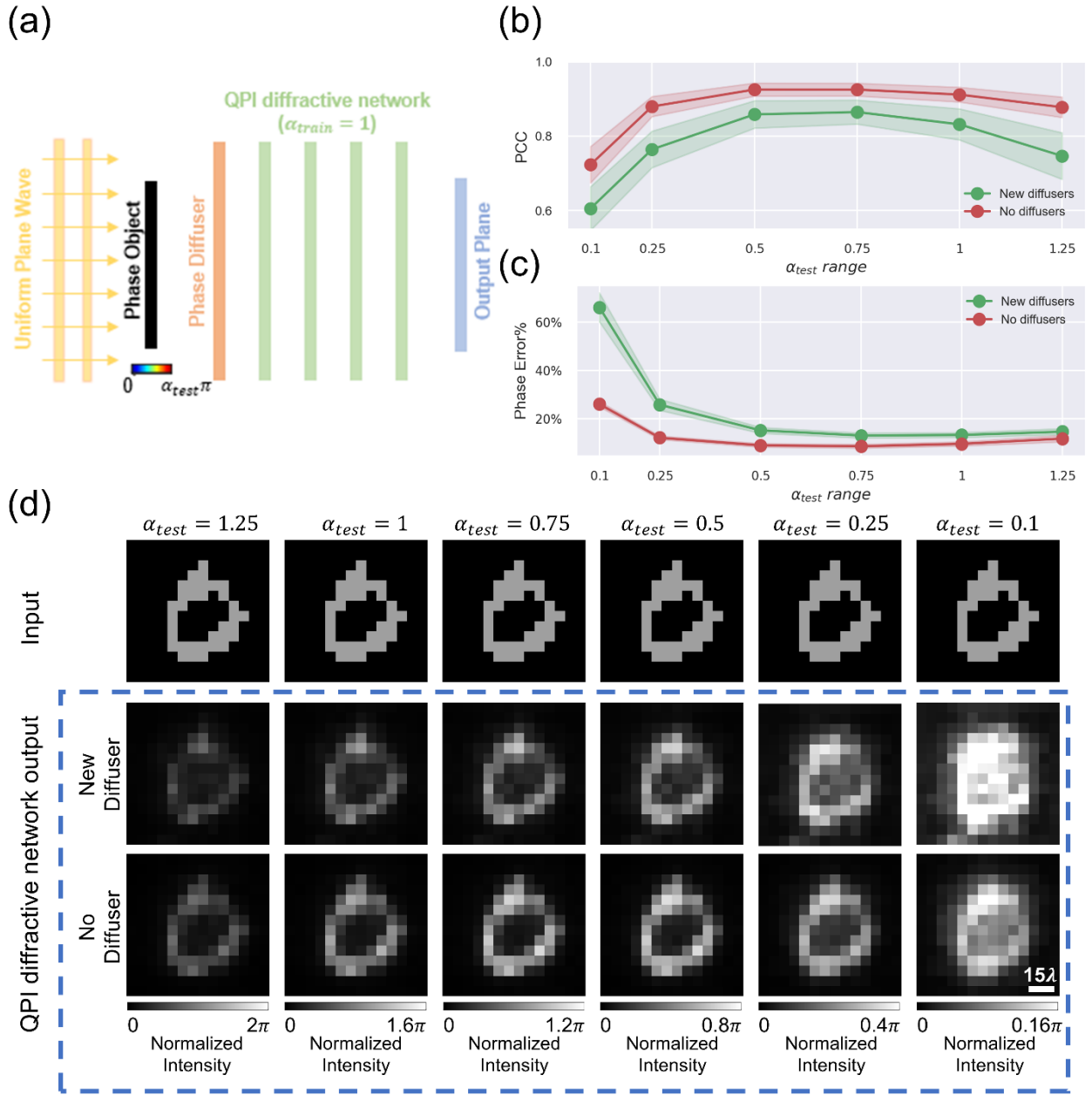

**Figure 4. The impact of the input phase contrast on the QPI D²NN signal quality.** (a) A schematic of the diffractive QPI network that was trained with $\alpha_{train} = 1$. (b) The PCC values and (c) the percent phase errors of the diffractive QPI signals with respect to the ground truth images as a function of $\alpha_{test}$. (d) Pairs of ground truth binary phase-encoded images (top row) and the results of the QPI D²NN imaging through new random diffusers (middle row) and no diffusers (bottom row) for different levels of phase encoding ranges, $\alpha_{test}$. $L = L_{train} = L_{test} = 14\lambda$.

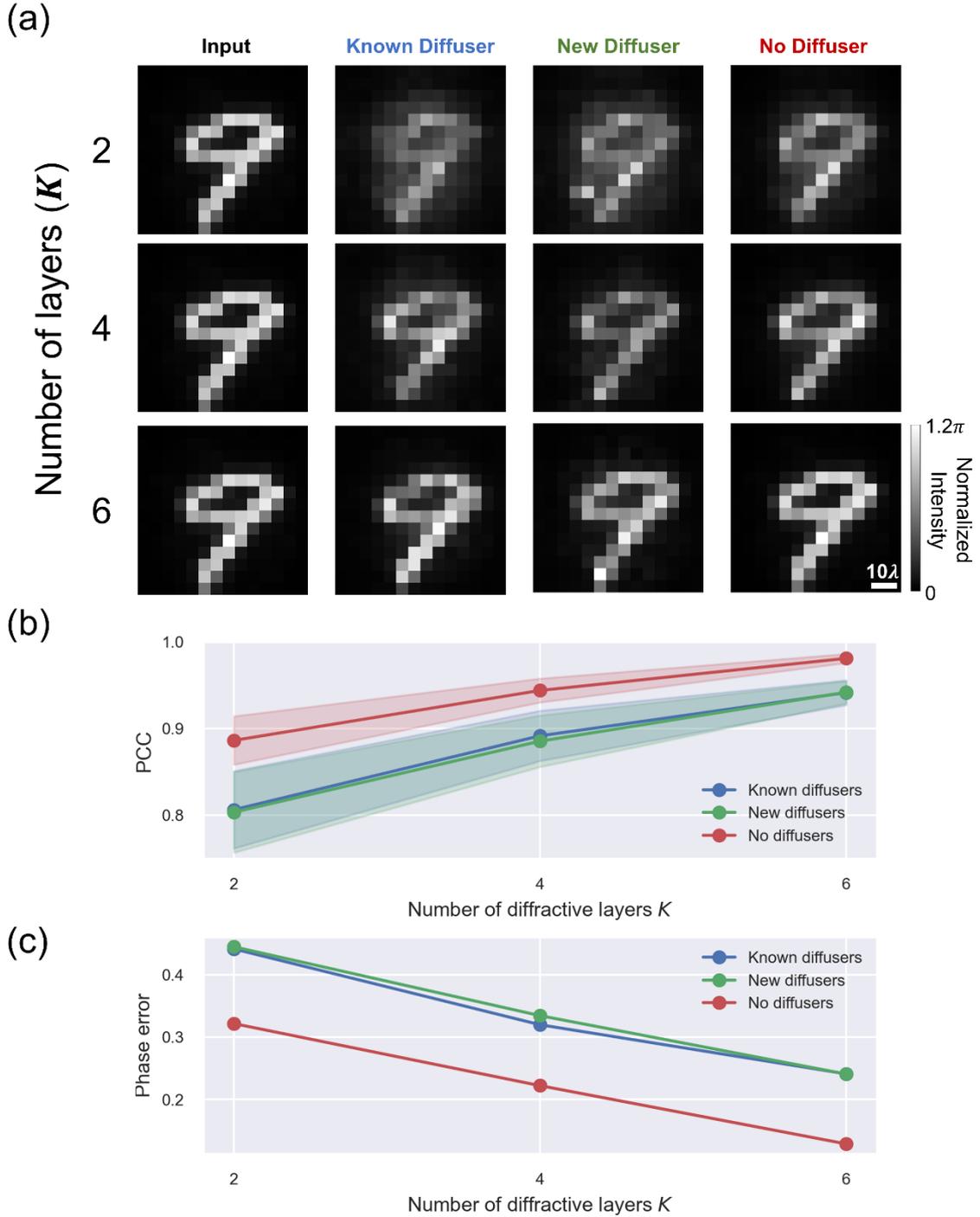

**Figure 5. Additional trainable diffractive layers improve the all-optical phase recovery performance for input phase objects imaged through random unknown diffusers. (a)** The output normalized intensity of the same phase object imaged by QPI D²NNs trained with $K = 2$, $K = 4$, and $K = 6$ diffractive layers through known (2nd column), new (3rd column) and no (4th column) diffusers. **(b)** The PCC and **(c)** the phase error values (in radians) of the QPI D²NN signals calculated with respect to the ground truth images as a function of the number of diffractive layers $K$. $L = L_{train} = L_{test} = 14\lambda$.

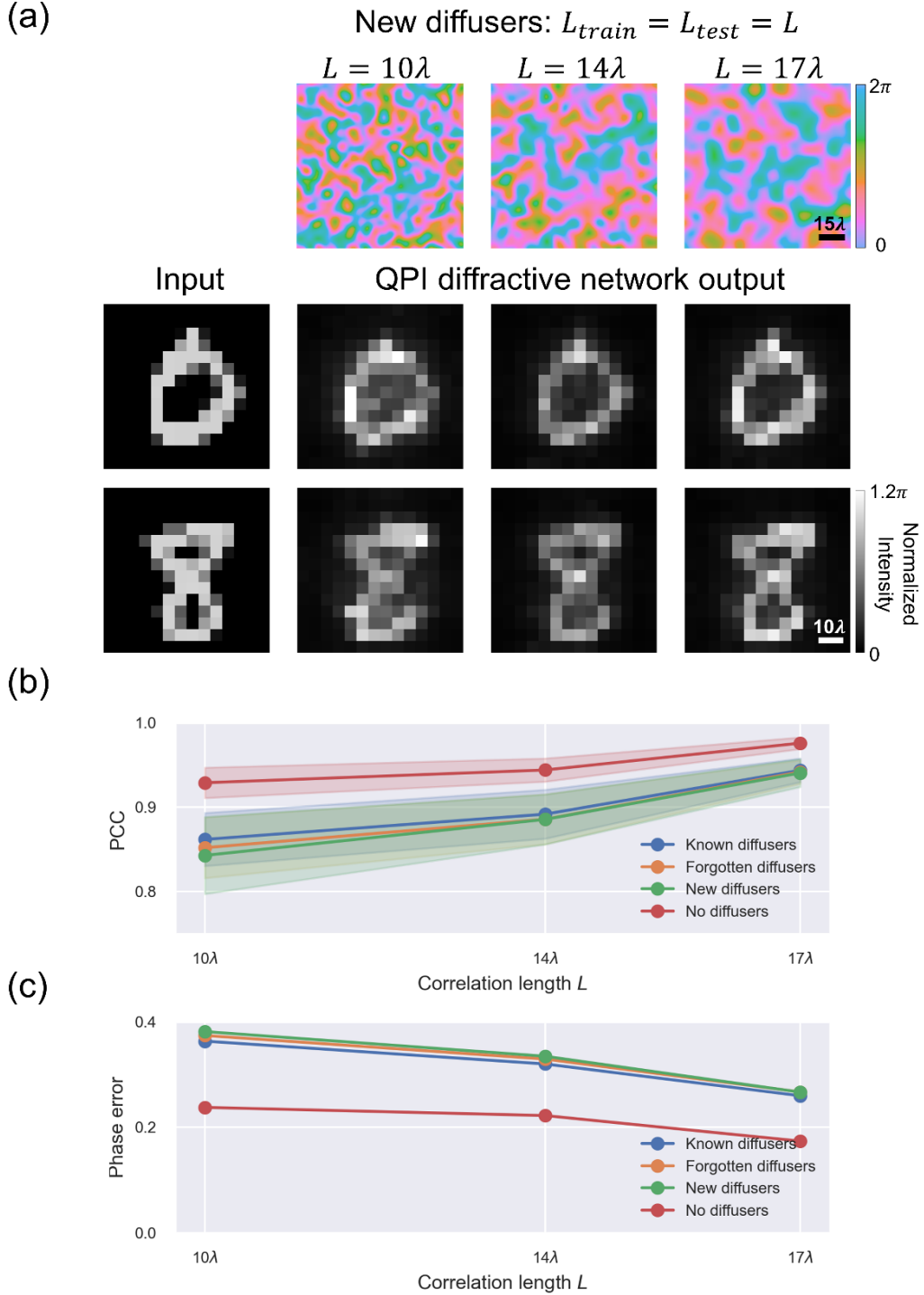

**Figure 6**. **Quantitative phase imaging through random unknown diffusers with different correlation lengths L.** (a) Output signal $I_{QPI}$ of the input phase objects imaged by diffractive QPI networks trained with $L_{train} = 10\lambda, 14\lambda$ and $17\lambda$ diffusers, seen through new random diffusers (1st row). (b) The PCC and (c) the phase error values (in radians) of the output signals of the QPI diffractive networks trained with $L_{train} = 10\lambda, 14\lambda$ and $17\lambda$, and tested through known, forgotten, new and no diffusers with the corresponding correlation length, i.e., $L_{test} = L_{train} = L$.

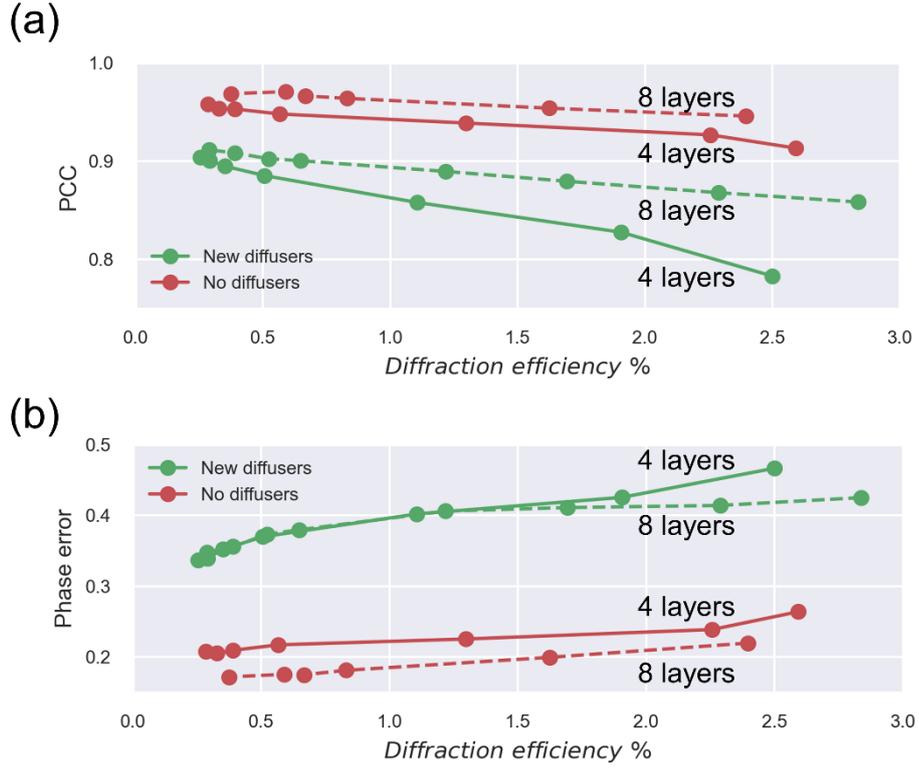

**Figure 7**. **The trade-off between diffractive QPI signal quality and the output power efficiency. (a)** The PCC values and **(b)** and the phase errors (in radians) of the QPI diffractive networks trained with various levels of diffraction efficiency penalty. These QPI D²NN models were trained using $\alpha_{train} = 1$ phase-encoded input samples selected from the MNIST dataset. Two D²NN setups using four and eight diffractive layers were trained and tested. $L = L_{train} = L_{test} = 14\lambda$.